\documentclass[preprint,12pt,3p]{elsarticle}

\usepackage{graphicx}
\usepackage{amssymb}

\usepackage{lineno}

\usepackage{algorithm} 
\usepackage{algpseudocode} 
\usepackage{color}

\usepackage{multirow}
\usepackage{caption}
\usepackage{subcaption}

\usepackage{amsmath}

\usepackage{hyperref}

\journal{arXiv}

\begin{document}

\begin{frontmatter}


\title{Micromobility Trip Origin and Destination Inference Using General Bikeshare Feed Specification (GBFS) Data}



\author[UF1]{Yiming Xu}
\author[UF1]{Xiang Yan}
\author[UAB]{Virginia P. Sisiopiku}
\author[FAU]{Louis A. Merlin}
\author[MS]{Fangzhou Xing}
\author[UF1]{Xilei Zhao\corref{cor1}}
\cortext[cor1]{Corresponding author. Address: 1949 Stadium Rd, Gainesville, FL 32611. Phone: +01 352-294-7159. Email: xilei.zhao@essie.ufl.edu.}

\address[UF1]{Department of Civil and Coastal Engineering, University of Florida, Gainesville, FL}
\address[UAB]{Department of Civil, Construction, and Environmental Engineering, University of Alabama at Birmingham, Birmingham, AL}
\address[FAU]{School of Urban and Regional Planning, Florida Atlantic University, Boca Raton, FL}
\address[MS]{Microsoft Corporation, One Microsoft Way, Redmond, WA}

\begin{abstract}
Emerging micromobility services (e.g., e-scooters) have a great potential to enhance urban mobility but more knowledge on their usage patterns is needed. The General Bikeshare Feed Specification (GBFS) data are a possible source for examining micromobility trip patterns, but efforts are needed to infer trips from the GBFS data. Existing trip inference methods are usually based upon the assumption that the vehicle ID of a micromobility option (e-scooter or e-bike) does not change, and so they cannot deal with data with vehicle IDs that change over time. In this study, we propose a comprehensive package of algorithms to infer trip origins and destinations from GBFS data with different types of vehicle ID. We implement the algorithms in Washington DC by analyzing one-week (last week of February 2020) of GBFS data published by six vendors, and we evaluate the inference accuracy of the proposed algorithms by R-squared, mean absolute error, and sum absolute error. We find that the R-squared measure is larger than 0.9 and the MAE measure is smaller than 2 when the algorithms are evaluated with a $400m\times400m$~grid, and the absolute errors are relatively larger in the downtown area. The accuracy of the trip-inference algorithms is sufficiently high for most practical applications. 
\end{abstract}

\begin{keyword}
Micromobility \sep e-scooters \sep GBFS data \sep trip origin and destination inference \sep Washington D.C. 


\end{keyword}

\end{frontmatter}


\section{Introduction}
\label{S:1}

In recent years, shared micromobility services are experiencing an explosive growth in urban areas \citep{nacto2019a,mckenzie2019spatiotemporal}. Micromobility refers to a mode of transportation that involves small, single-passenger modes rented for short term use, such as e-scooters, docked bikes, and dockless bikes. Among all the micromobility options, e-scooters are growing at the fastest pace, largely owing to their flexibility, convenience, affordability, and the fun factor. They are especially attractive for travelers for short-distance trips. Micromobility also offers a potential solution to the ``first-mile/last-mile'' problem (the problem of public transit being unable to get passengers to the doorstep of their destinations) that has long troubled public transit. Given the potential benefits that e-scooters offer for urban mobility, a great interest exists among transportation professionals for understanding the usage patterns of micromobility services.

As a part of the micromobility permit requirement, cities often require micromobility providers to share data through application programming interfaces (APIs) that are prescribed by standard formats, including the General Bikeshare Feed Specification (GBFS) and the Mobility Data Specification (MDS). GBFS was initially developed as the open data standard for bike share system availability back in 2015, but now it is applicable for all shared micromobility systems \citep{nabsa2020github}. GBFS APIs report real-time information about available vehicles, which typically includes vehicle location, vehicle type (bike or scooter), and battery level. Created by the Los Angeles Department of Transportation (LADOT) in 2018, MDS extends GBFS to require additional information from mobility providers. The additional information may include data on unavailable vehicles in the network, trip characteristics, and trip trajectories \citep{regina2019medium}. However, the MDS has received limited adoption so far, and the MDS APIs are usually not made available to the public. Accordingly, we focus on GBFS in this study.

Some recent studies that examine the spatiotemporal patterns of scooter usage extracted trip information from GBFS data  \citep{mckenzie2019spatiotemporal,zou2020exploratory}. The trip inference method used in these studies usually assumes that the vehicle ID of e-scooters, as reported by GBFS APIs, does not change over time. In reality, however, many micromobility providers operate GBFS APIs that report dynamic vehicle IDs (i.e., the vehicle ID of a given scooter changes over time). This means that the application of the existing methods are limited to GBFS APIs that report static vehicle IDs, which is becoming less popular as micromobility providers switch to reporting dynamic vehicle IDs; many providers are switching due to privacy concerns, because dynamic vehicle IDs enhance rider privacy. Therefore, a systematic method to infer e-scooter trip data for different vehicle ID types is needed for the public to infer trips from GBFS data. 

To this end, this study develops a package of algorithms to infer trip origins and destinations for all existing vehicle ID types based on GBFS data provided by various vendors. We use three metrics including R-squared ($R^2$), mean absolute error ($MAE$), and sum absolute error ($SAE$) to validate the inference accuracy of proposed algorithms. All proposed algorithms have an R-squared larger than 0.9 when scooter location data is aggregated to a $400m\times400m$~grid, which is sufficiently high for most practical purposes. The inferred e-scooter trip origins and destinations show usage patterns that are consistent with previous studies \citep{mckenzie2020urban,mckenzie2019spatiotemporal}. Researchers and public agencies can use our package to estimate the trip information for micromobility services with publicly accessible APIs (e.g., GBFS), and then they can derive useful insights to facilitate city planning and decision-making relative to e-scooters.

The remainder of the paper is structured as follows. The next section presents a literature review on previous studies on e-scooter services. Section 3 describes the data collection procedures and the proposed algorithms in detail. Section 4 presents an application of the trip-inference algorithms in Washington D.C., following with a validation. Section 5 concludes the paper by summarizing findings, identifying limitations, and suggesting future work.

\section{Literature Review}

Studies on e-scooter usage patterns are growing but still limited in supply. Some studies analyze trip data published on a city government's open data portal \citep{bai2020dockless,caspi2020spatial}. This kind of data directly provides information (e.g., trip origin, trip destination, trip duration, and trip distance) on every trip. For example, \citet{bai2020dockless} explored e-scooter ridership patterns and analyzed the relationship between e-scooter usage and the built environment in Austin, TX and Minneapolis, MN, based on the shared micromobility vehicle trips data published by the City of Austin and the City of Minneapolis \citep{austin2020trip,minn2020trip}. However, e-scooter trip data are unavailable to the public for most cities, because these cities have not mandated trip-level data sharing for their micromobility providers. Some authors thus acquired the e-scooter trip data from the city via personal connections. For example, \citet{liu2019analysis} used the data provided from the City of Indianapolis to analyze the spatiotemporal patterns of e-scooter trips. 

Other studies scraped the publicly accessible GBFS or data published by the micromobility vendors and then processed the data to derive needed information (e.g., trip origin and destination, trip trajectory). For example, \citet{mckenzie2019spatiotemporal} compared spatial and temporal patterns of e-scooter-share and bike-share usage in Washington, D.C. using GBFS data. In another study, \citet{mckenzie2020urban} further explored spatial and temporal differences in usage patterns between six vendors in Washington, D.C., including Bird, Lime, Lyft, Skip, Spin, and Jump. The methods used in the two McKenzie studies to derive e-scooter trips from scraped GBFS data were the same: a trip was identified as the time and location of when an e-scooter last appeared available in the system, to the time and location of when the same e-scooter next appeared available in the set of scrapped data. However, since the vehicle ID was used to identify trips, this trip inference method requires the vehicle ID of e-scooter to be consistent over time. \citet{zou2020exploratory} examined e-scooter travel patterns at the street-segment level based on e-scooter trip trajectories in Washington, D.C. E-scooter trip trajectories were pinpointed based on the attribute in the data that indicated whether an e-scooter was in use. However, this method requires static vehicle ID as well as information on all e-scooters in the system. Only one vendor's data met these requirement and was used in this study; this kind of data is no longer available in Washington, DC now. \citet{baltra2020data} discussed the privacy issues of GBFS specification and used GBFS data in Los Angeles to infer trips. The logic of trip inference in \citet{baltra2020data} was the same as \citet{mckenzie2019spatiotemporal} and \citet{mckenzie2020urban} and relied upon a static scooter ID. \citet{zhu2020understanding} scraped the docked e-scooter and station status data in bike-sharing systems of Singapore. Based on the data, they investigated spatio-temporal heterogeneity of bike-sharing and e-scooter-sharing systems in two urban areas in Singapore. E-scooter trips were identified by continuously checking the vehicle ID of e-scooters in each station. If a vehicle ID disappeared, this was considered a trip origin, and if a vehicle ID appeared, this was considered a trip destination. A summary of existing studies based on inferred e-scooter trips is presented in Table~\ref{tab:lit}. In conclusion, the existing e-scooter trip inference methods can infer trips when providers use a static vehicle ID, but cannot deal with dockless e-scooter data with dynamic vehicle ID.

\begin{table}[!ht]
\caption{Summary of existing studies based on inferred e-scooter trips}\label{tab:lit}
\begin{center}
\resizebox{1\textwidth}{!}{
\begin{tabular}{|l|l|l|l|}
\hline
Study        & Data source                                                   & Inference method                                                                                                             & \begin{tabular}[c]{@{}l@{}}Reliance on \\ static ID\end{tabular} \\ \hline
\citet{mckenzie2019spatiotemporal} & GBFS                                                          & \begin{tabular}[c]{@{}l@{}}Disappearance and re-appearance \\ of the vehicle in the dataset\end{tabular}                     & Yes                                                                  \\ \hline
\citet{mckenzie2020urban} & GBFS                                                          & \begin{tabular}[c]{@{}l@{}}Disappearance and re-appearance\\ of the vehicle in the dataset\end{tabular}                      & Yes                                                                  \\ \hline
\citet{zou2020exploratory}     & GBFS                                                          & \begin{tabular}[c]{@{}l@{}}Checking the attribute in the data that \\ indicated whether an e-scooter was in use\end{tabular} & Yes                                                                  \\ \hline
\citet{baltra2020data}  & GBFS                                                          & \begin{tabular}[c]{@{}l@{}}Disappearance and re-appearance\\  of the vehicle in the dataset\end{tabular}                     & Yes                                                                  \\ \hline
\citet{zhu2020understanding}     & \begin{tabular}[c]{@{}l@{}}Bike-sharing\\ system\end{tabular} & \begin{tabular}[c]{@{}l@{}}Checking the vehicle ID \\ of e-scooters in each station\end{tabular}                             & Yes                                                                  \\ \hline
\end{tabular}
}
\end{center}
\end{table}

However, micromobility providers are moving away from static vehicle IDs out of the concern for privacy issues. A recent effort to preserve user privacy was to adopt dynamic vehicle ID in their published GBFS data, which is a growing trend since the release of GBFS Version 2\footnote{Improvements include micromobility provider integration into mobility apps and stronger rider privacy.}\citep{nabsa2020github}. For example, Lime published static vehicle IDs before September 24, 2019 but switched to dynamic IDs afterwards. The existing trip-inference methods will not work for these kinds of data. Since vendors concentrate their vehicles at different locations, which results in distinctive spatio-temporal usage patterns across vendors \citep{mckenzie2020urban}, omitting trip data from vendors that publish dynamic vehicle IDs would provide an incomplete picture of e-scooter demand in the city. Trip-inference \footnote{Trip inference used in the paper is not only referring to trips per se, instead, we are looking at inferring separate trip origins and destinations.} approaches that can infer trips from GBFS data with different vehicle ID types are needed for a complete understanding of e-scooter-related travel behavior.

\section{Methods}
This section describes data collection and three algorithms we have developed to infer trip origins and destinations using GBFS data.

\subsection{Data Collection}

The GBFS specification requires that vendors publish certain data feeds as JSON files, which are usually made accessible via public APIs provided by each vendor. Among these data feeds, the one that can be used to infer trip origins and destinations is \textit{free\_bike\_status}, which provides location information of all currently available vehicles in the system. The attributes in \textit{free\_bike\_status} are listed in Table~\ref{tab:fbs}. A data sample is as follows:
\begin{small}
\begin{verbatim}
{"last_updated": 1582528501, "ttl": 300, "data": {"bikes": [{"bike_id":8982,
"lat":38.8962,"lon":-76.9592,"is_reserved":0,"is_disabled":0},{"bike_id":9408,
"lat":38.8797,"lon":-77.0100,"is_reserved":0, "is_disabled":0}]}}
\end{verbatim}
\end{small}

\begin{table}[!ht]
	\caption{Attributes in \textit{free\_bike\_status}}
	\label{tab:fbs}
	\small
	\begin{center}
		\begin{tabular}{l l l}\hline
			Attribute & Description & Type \\
			\hline
			\textit{last\_updated} & POSIX timestamp indicating the last time the data in was updated & Integer \\
			\textit{ttl}   & Seconds before the data in this feed will be updated again  & Integer \\
			\textit{bike\_id}   & Unique identifier of an e-scooter & String \\
			\textit{lat}   & Latitude of the e-scooter location & Number \\
			\textit{lon}   & Longitude of the e-scooter location & Number \\
			\textit{is\_reserved}\footnotemark    & Is the e-scooter currently reserved for someone else & Integer \\
			\textit{is\_disabled} & Is the e-scooter currently disabled (broken) & Integer \\
			\hline
		\end{tabular}
	\end{center}
\end{table}
\footnotetext{Since only the information on available vehicles is published, this attribute is always 0.}

The data feeds are updated every \textit{time to live (TTL)} seconds. Since no specific value is required by the GBFS specification, the \textit{TTL} varies among vendors. Typical \textit{TTL} in GBFS includes 0\footnote{0 means the data should always be refreshed.}, 300, and 600. In this study, for the data feeds with a \textit{TTL} below 60 seconds, we scrape the data every minute; otherwise, we use the \textit{TTL} seconds as scraping interval.

\subsection{Trip Origins and Destinations Inference Algorithms}
As discussed in previous section, the GBFS data only provide information on currently available e-scooters in the system. In other words, information on e-scooters in use is not available. In addition, the vendors may adopt different e-scooter ID generating strategies. Currently, there are three different e-scooter ID generating strategies in GBFS data:
\begin{enumerate}
  \item \textit{Static Vehicle ID}: Vehicle ID does not change until the e-scooter is taken out of service.
  \item \textit{Resetting Vehicle ID}: Vehicle ID of corresponding e-scooter randomizes after every trip, but is otherwise static. For example, when a user terminates a trip and returns the e-scooter, the system will assign the e-scooter a new random vehicle ID.
  \item \textit{Dynamic Vehicle ID} : Vehicle ID for all e-scooters in the system randomizes every several minutes\footnote{Typical time intervals: 30 minutes, 1 hour. Time intervals vary among different vendors.}. 
\end{enumerate}
The three ID generating strategies produce three different types of e-scooter vehicle ID. The e-scooter trip origins and destinations inference logic differs for different ID types. Therefore, for each e-scooter vehicle ID type, we develop an algorithm to infer e-scooter trip origins and destinations from GBFS data. The three algorithms are summarized in Table~\ref{tab:sal}.

\begin{table}[!ht]
\caption{Summary of Inference Algorithms}\label{tab:sal}
\centering
\resizebox{1\textwidth}{!}{
\begin{tabular}{|l|l|l|l|l}
\cline{1-4}
\textbf{Algorithm} & \textbf{Application Scenario}                                        & \textbf{Output}                                                                    & \textbf{Causes of errors}                                                                                                                               &  \\ \cline{1-4}
Algorithm 1        & Static Vehicle ID                                                & OD pairs                                                                          & GPS error, data update frequency.                                                                                                                       &  \\ \cline{1-4}
Algorithm 2        & \begin{tabular}[c]{@{}l@{}}Resetting Vehicle ID \end{tabular} & \begin{tabular}[c]{@{}l@{}}Unlinked trip origins \\ and destinations\end{tabular} & \begin{tabular}[c]{@{}l@{}}GPS error, data update frequency,  \\ launch and elimination of vehicles.\end{tabular}                                       &  \\ \cline{1-4}
Algorithm 3        & \begin{tabular}[c]{@{}l@{}}Dynamic Vehicle ID \end{tabular} & \begin{tabular}[c]{@{}l@{}}Unlinked trip origins \\ and destinations\end{tabular} & \begin{tabular}[c]{@{}l@{}}GPS error, data update frequency,  \\ launch and elimination of vehicles, \\ movement of e-scooters not in use.\end{tabular} &  \\ \cline{1-4}
\end{tabular}
}
\end{table}

Vendors sometime relocate e-scooters away from locations where they are not being used to high demand areas. We call this type of trip a ``rebalancing trip'' in this paper. To recharge low-battery e-scooters in time, some vendors (e.g. Lime) pays gig workers to recharge e-scooters at their residence. Participants are instructed to pick up e-scooters with low batteries, and drop them off to specific locations after recharging. This type of trip is named a ``juicing trip'' \citep{mckenzie2019spatiotemporal}. Both of rebalancing trip and juicing trip are not real trips conducted by travelers, thus our intention is to exclude such trips.

\subsubsection{Static Vehicle ID: Algorithm 1}
Since the GBFS data only provides real-time information on available e-scooters, once a user unlocks an e-scooter and starts a trip, information on this e-scooter disappears from the corresponding data feed. When the trip terminates, information on the e-scooter will reappear in the data feed. Therefore, we can infer that information on an e-scooter disappearing in the data feeds indicates a trip origin of this e-scooter. Similarly, information on an e-scooter reappearing in the data feeds indicates a trip destination. The inference logic is similar to previous studies \citep{mckenzie2019spatiotemporal,minn2020trip,baltra2020data}. Based on this logic, we develop an algorithm to infer trip origins and destinations for e-scooters with \textit{Static Vehicle ID}. The algorithm is presented in detail in Algorithm~\ref{alg1}.

\begin{algorithm}[!h]
	\caption{OD Pair Inference for Data with \textit{Static Vehicle ID}}\label{alg1}
	\begin{algorithmic}[1]
	    \State \textbf{input} all e-scooter records $D_{all}$, data scraping interval $\Delta t_s$
	    \State $V \leftarrow$ A set of unique Vehicle IDs in $D_{all}$
		\For {$j$ in $1:length(V)$}
		\State $D \leftarrow$ Records with vehicle ID = $V_j$ in $D_{all}$
		\State $D_s \leftarrow$ Sort records in $D$ by time 
		\State $i\leftarrow1$
		\State $n\leftarrow$ Number of records in $D_s$
		\While {$i<n$}
		    \State Calculate time interval $\Delta t_i$ between $i$th and $(i+1)$th records
		    \If {$\Delta t_i>\Delta t_s$}
		    \State Trip origin $Ori \leftarrow i$th record 
		    \State Trip destination $Des \leftarrow (i+1)$th record
		    \EndIf
		    \State \textbf{output} An OD pair $(Ori, Des)$
		    \State $i\leftarrow i+1$
		\EndWhile
		\EndFor
	\end{algorithmic} 
\end{algorithm} 

When e-scooter vendors publish valid data with a high update frequency, this algorithm can produce accurate inference results on OD locations. Theoretically, the algorithm can identify all actual trips if data feeds are updated frequently enough, i.e. every second. In addition, since the e-scooter vehicle ID is static, this algorithm can infer trip OD pairs. Inference error is possible if there is GPS error or a low data update frequency. Launch and elimination of vehicles will not result in inference errors. The algorithm only infers linked OD pairs, thus the first record will not be identified as trip destination and the last record will not be identified as trip origin. When a e-scooter is in a rebalancing trip or juicing trip, since the e-scooter is locked, some vendors may mark this e-scooter as "available" and publish information on it to \textit{free\_bike\_status}. In this case, we can observe movement of e-scooters in the data feeds but the algorithm will not categorize these trips as real trips conducted by travelers. 

The analyst may seek to exclude rebalancing and juicing trips by considering additional criteria such as travel distance, trip speed, or time of day \cite{mckenzie2019spatiotemporal, zou2020exploratory}. In this study, we use the same criteria as in \citet{mckenzie2019spatiotemporal}. Trips that lasted longer than 2 hours, trips with average speeds greater than 15 miles per hour, and trips with average speeds lower than 2.2 miles per hour are removed in this study. 

\subsubsection{Resetting Vehicle ID}
When an e-scooter using \textit{Resetting Vehicle ID} starts a trip, information on the e-scooter will disappear in the data feed. After the trip terminates, information on this e-scooter will reappear but with a different ID. The ID is randomly generated so we cannot match the new ID with previous one. Therefore, we cannot infer OD pairs using this kind of data. But we can infer trip origins and destinations separately. The e-scooter starts using a new ID when a trip terminates. When next trip starts, this ID then disappears in the data feeds. Therefore, when an e-scooter appears in the system with an ID, we mark this as a trip destination; then at a later time point, when the e-scooter with this same ID disappears in the system, we mark this as a trip origin. In other words, for all records with a unique vehicle ID, the earliest record indicates a trip destination and the latest record indicates the origin of next trip. Based on this logic, an algorithm to infer trip origins and destinations for e-scooters with \textit{Resetting Vehicle ID} is developed. The algorithm is presented in Algorithm~\ref{alg2}.

\begin{algorithm}[!h]
	\caption{Origin and Destination Inference for Data with \textit{Resetting Vehicle ID}}\label{alg2}
	\begin{algorithmic}[1]
	    \State \textbf{input} all e-scooter records $D_{all}$
	    \State $V \leftarrow$ A set of unique Vehicle IDs in $D_{all}$
		\For  {$j$ in $1:length(V)$}
		\State $D \leftarrow$ Records with vehicle ID $=V_j$ in $D_{all}$
		\State $D_s \leftarrow$ Sort records in $D$ by time
		\State A trip destination $Des \leftarrow$ the first record in $D_s$
		\State \textbf{output} $Des$
		\State A trip origin $Ori \leftarrow$ the last record in $D_s$
        \State \textbf{output} $Ori$
		\EndFor
	\end{algorithmic} 
\end{algorithm} 

Like Algorithm 1, the inference accuracy of this algorithm is related to data update frequency and GPS error. In addition, this algorithm may overestimate trip origins and destinations. If a new e-scooter is added to the system, the first record of this e-scooter will also be incorrectly identified as a trip destination. Similarly, when a broken e-scooter is removed from the system, the last record of this e-scooter is incorrectly identified as a trip origin. The sources of error are summarized in Table~\ref{tab:sal}. The adverse effects of these limitation are evaluated in inference accuracy section that follows. 

\subsubsection{Dynamic Vehicle ID}

Scooters with \textit{Dynamic Vehicle ID} change vehicle ID periodically, according to a certain time interval (e.g. 30 minutes). This ID generating strategy is a good way to protect user privacy but makes it challenging to infer trip origins and destinations. However, it is still possible to infer trip origins and destinations using our proposed method.

The first step is to create a series of zones and count the total number of scooters in each zone at regular intervals. Here we use A and B to denote specific zones. The zones should be large enough that GPS location error between time periods for the same stationary scooter does not lead to an inference of a false trip.


Consider a series of consecutive time steps $[T, T+1, ..., T+K]$, for $K>1$. Suppose an e-scooter trip starts from location $A$ at $T$ and ends at location $B$ at $T+K$. When we compare locations of all the e-scooters at $T$ and $T+1$, we can find that this e-scooter disappears at location $A$. Similarly, when we compare locations of e-scooters at $T+K-1$ and $T+K$, we can find an e-scooter appears in location $B$. Based upon this logic, we can develop an algorithm to infer trip origins and destinations. Since the time interval of ID updates is usually larger than the data scraping interval, we also use vehicle ID information to enhance inference accuracy. For example, if a vehicle ID is found both in data feeds at $T$ and $T+1$, we can conclude that the e-scooter with this ID made no trip during this time interval $[T,T+1]$. The algorithm to infer trip origins and destinations for data with \textit{Dynamic Vehicle ID} is presented in Algorithm~\ref{alg3}.

\begin{algorithm}[ht!]
	\caption{Origin and Destination Inference for Data with \textit{Dynamic Vehicle ID}}\label{alg3}
	\begin{algorithmic}[1]
	    \State \textbf{input} all e-scooter records $D_{all}$, buffer $b$ \footnotemark
		\For {Every time interval $[T,T+1]$}
		\State $D_T \leftarrow$ Records at $T$ in $D_{all}$
		\State $D_{T+1} \leftarrow$ Records at $T+1$ in $D_{all}$
		\State $V \leftarrow$ A set of unique Vehicle IDs in $D_T$
		\For {$k$ in $1:length(V)$}
		\If {$V_k$ in $D_{T+1}$}
		    \State Remove records with vehicle ID $=V_k$ in $D_{T}$ and $D_{T+1}$
		\EndIf
		\EndFor
		\\
		\Repeat
		\For{Every record $i$ in $D_{T}$}
		\For{Every record $j$ in $D_{T+1}$}
		\State Calculate Euclidean distance $d_{ij}$ between $i$ and $j$
		\EndFor
		\EndFor
		\State $d_{mn}\leftarrow$ minimum distance in $d_{ij}$
		\State $(m,n)\leftarrow $ index of $d_{mn}$
		\State Remove record $m$ in $D_{T}$ 
		\State Remove record $n$ in $D_{T+1}$
		\Until{$d_{mn}>b$}
		\\
		\State Trip origin $Ori \leftarrow$ records in $D_T$
		\State \textbf{output} $Ori$
		\State Trip destination $Des \leftarrow$ records in $D_{T+1}$
		\State \textbf{output} $Des$
		\EndFor
	\end{algorithmic} 
\end{algorithm} 
\footnotetext[6]{Minimum distance threshold to distinguish a trip from GPS error. We use $b=100m$ in case study.}

In the case of Algorithm 3, there are more potential sources of error. Inference error may result from GPS error, low data update frequency, and the launch or elimination of e-scooters. Besides, movements of e-scooters which are not real trips (e.g., rebalancing and juicing trips) may be recognized as trips by the algorithm, which may cause extra inference errors. The sources of error are summarized in Table~\ref{tab:sal}. The inference accuracy of this algorithm is discussed in Subsection 4.2.

\section{Application and validation of the proposed algorithms}
We collect published GBFS data of six e-scooter services vendors\footnote{We also collected GBFS data of Razor. However, the data update frequency of Razor is 1 hour, which is too long for trip inference. Therefore, we did not use these data in this study.} (i.e., Bird, Jump, Lime, Lyft, Skip, and Spin) in Washington D.C. from February 24, 2020 to March 01, 2020. As described in the data collection section, data feeds are scraped using API\footnote{https://ddot.dc.gov/page/dockless-api} provided by the vendors. Since the data update frequency varies among vendors, we use different scraping intervals. Specifically, scraping interval for Bird, Jump, Skip, and Spin data is 60 seconds, and scraping interval for Lime and Lyft is 300 seconds. For vehicle ID generating strategy, Jump, Skip and Spin use \textit{Static Vehicle ID}. Bird uses \textit{Resetting Vehicle ID}. Lime and Lyft use \textit{Dynamic Vehicle ID}. After data collection, trip origins and destinations are inferred by the corresponding algorithm.

\subsection{Inference Results}
\begin{figure}
     \centering
     \begin{subfigure}[h]{1\textwidth}
         \centering
         \includegraphics[width=\textwidth]{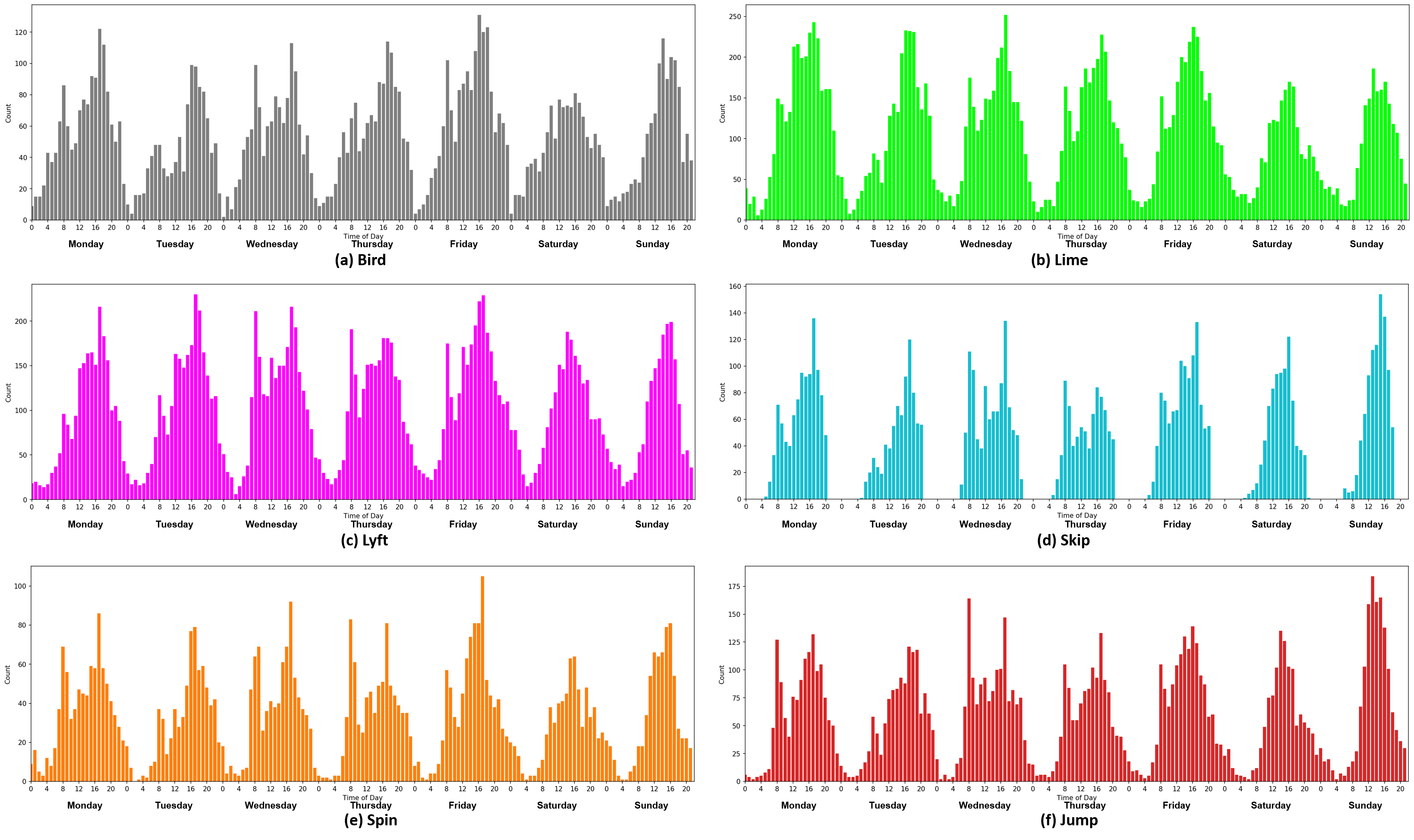}
         \caption{Trip Origins}
         \label{fig:t_ori}
     \end{subfigure}
     \hfill
     \begin{subfigure}[h]{1\textwidth}
         \centering
         \includegraphics[width=\textwidth]{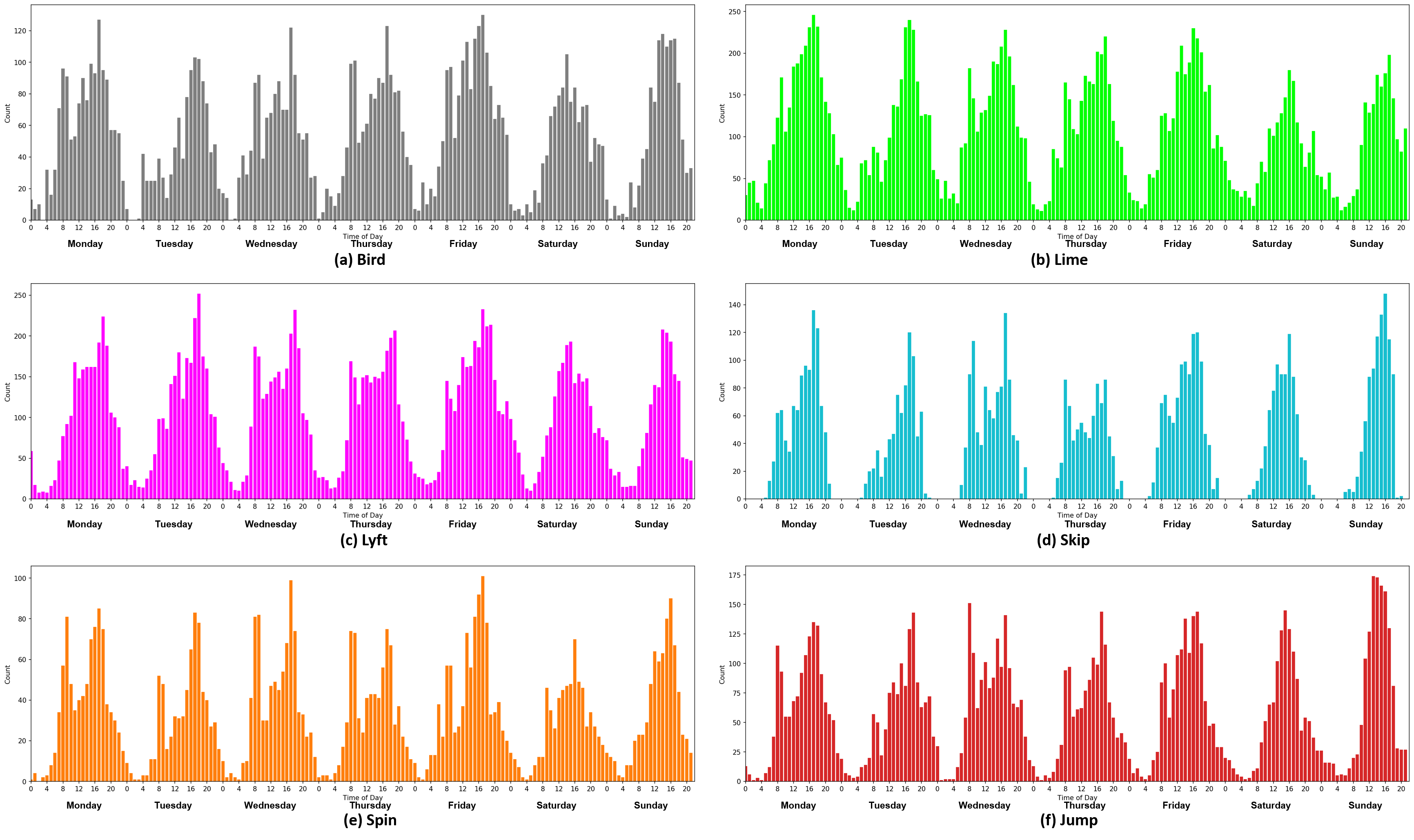}
         \caption{Trip Destinations}
         \label{fig:t_des}
     \end{subfigure}
        \caption{Temporal distributions of trip origins and destinations for different vendors in Washington, D.C.}
        \label{fig:temporal}
\end{figure}

Based on the data collected, 65,601 trip origins and 65,600 trip destinations are inferred in total. The temporal distribution of trip origins and destinations is presented in Figure~\ref{fig:temporal}. Note that each y-axis is different as the trip volume differs among vendors. Histogram colors are based on the dominant colors of the service logo.

According to Figure~\ref{fig:temporal}, while the six vendors have different trip volumes, the temporal usage patterns are generally similar, which is consistent with conclusions in \citet{mckenzie2020urban}. All of the six vendors have two significant trip peaks during weekdays. The morning peak hours are 8:00 a.m. to 10:00 a.m., and the afternoon peak hours are 5:00 p.m. to 7:00 p.m. But on weekends, there is only one peak, during the afternoon from 3:00 p.m. to 5:00 p.m. This difference between weekdays and weekends presumably results from commuting trips, which  are more common during weekdays. Compared with the study by \citet{mckenzie2020urban}, where data from December 2018 to March 2019 were used, the morning and afternoon peaks in weekdays are more prominent in our study. This result suggests that the proportion of commuting trips in e-scooter trips is increasing over time. Since the travel times of most commuter e-scooter trips are less than one hour and trips are aggregated by hour of the day, there are no significant differences between the temporal distributions of trip origins and destinations. But we do observe that the count of trip destinations is larger than count of trip origins during the second half of morning peak (i.e., 9:00 a.m. to 10:00 a.m.). This phenomenon may be caused by commuting trips too, most of which start from 8:00 a.m. to 9:00 a.m. and end before 10:00 a.m.

\begin{figure}
     \centering
     \begin{subfigure}[h]{1\textwidth}
         \centering
         \includegraphics[width=\textwidth]{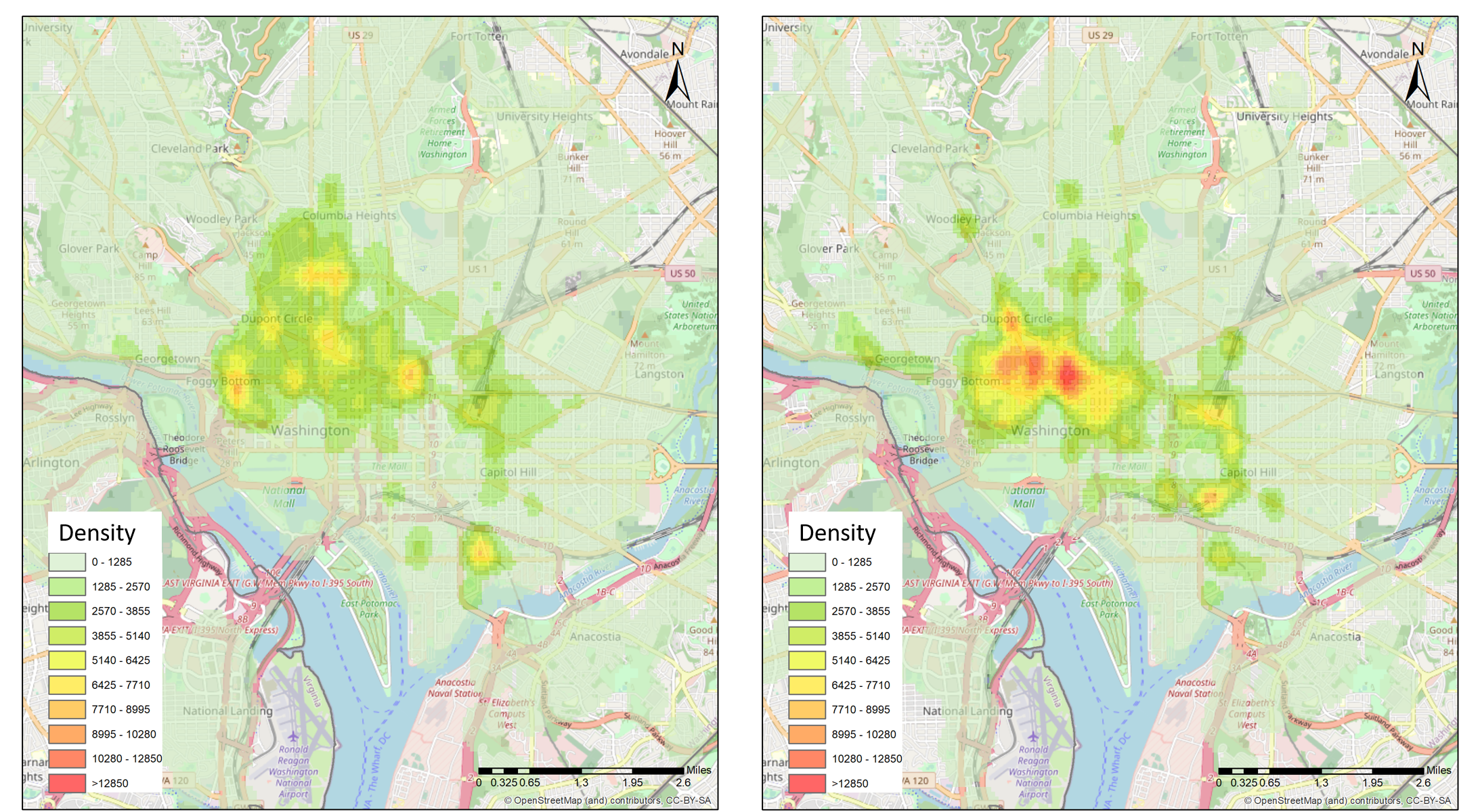}
         \caption{Morning Peak (Left: trip origin; Right: trip destination.)}
         \label{fig:mor_peak}
     \end{subfigure}
     \hfill
     \begin{subfigure}[h]{1\textwidth}
         \centering
         \includegraphics[width=\textwidth]{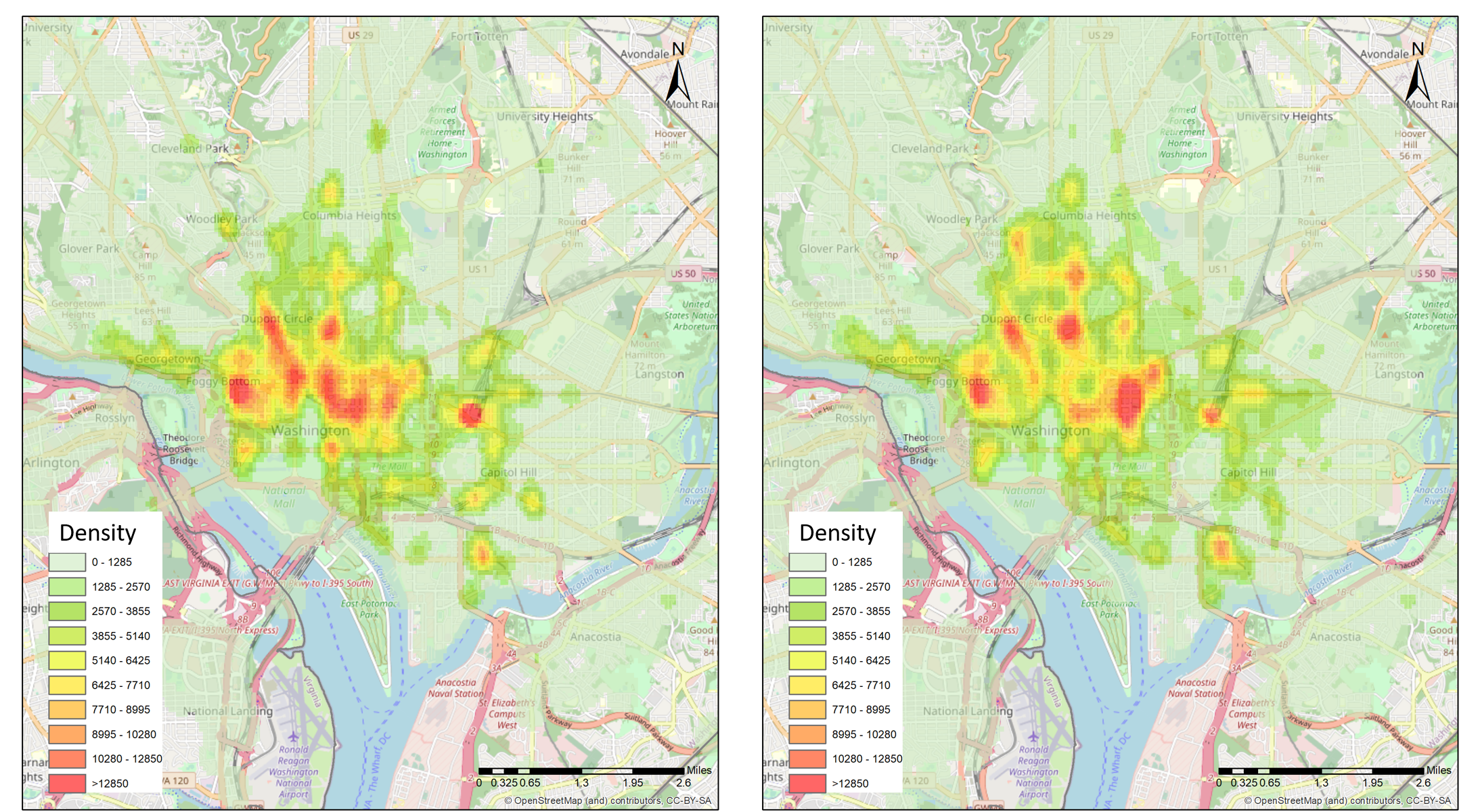}
         \caption{Afternoon Peak (Left: trip origin; Right: trip destination.)}
         \label{fig:aft_peak}
     \end{subfigure}
        \caption{Spatial distributions of trip origins and destinations in Washington, D.C. (Density unit: trips per day per~\textbf{$km^2$}.)}
        \label{fig:spatial}
\end{figure}

Using the data we then generate heat maps based on trip origin and destination density to explore spatial e-scooter usage patterns during peak hours. The results are presented in Figure~\ref{fig:spatial}. The general spatial pattern of peak-hour trips is that the trip density is significantly high in downtown areas, especially along the arterials. The typical hot spots of e-scooter usage include Dupont Circle, Foggy Bottom, Mount Vernon Square, Union Station and the U.S. Department of Transportation. Interestingly, all of these areas are centered or close to metro stations. This may indicate that the e-scooter services provide connections between transit stations and origins or destinations of commuters. If so, e-scooter services may help address the ``first-mile/last-mile'' problem. For morning peaks, the trip destinations are more concentrated than trip origins. This could be because most of the morning commuting trips are from home to work, and the work places are more likely to congregate in downtown areas. For afternoon peaks, there is a significant hot spot for trip destinations in the Penn Quarter, which is a popular entertainment district with bars, restaurants, and shopping. This pattern makes sense because people may use e-scooters to travel to leisure activities after work.

\subsection{Evaluation of the Proposed Algorithms}
GPS error and the error caused by data update frequency are systematic and irreducible. With the same data set, the two errors have the same effect on inference accuracy for the three algorithms. As discussed in the methods section, the inference error of Algorithm~\ref{alg1} only results from GPS error and the error caused by data update frequency. Theoretically, Algorithm~\ref{alg1} is expected to have the best performance among the three algorithms. Therefore, we use the inference result of Algorithm~\ref{alg1} as the \textit{benchmark} to evaluate the performance of Algorithms 2 and 3.

The inference accuracy of proposed algorithms is evaluated by three metrics: R-squared $(R^2)$, Mean Absolute Error $(MAE)$, and Sum Absolute Error $(SAE)$. We firstly divide the research area into a regular square grid with zones of equal size. Then we aggregate the trip origins and destinations into these grids based on location and then count the number of origins and destinations in each grid cell. Since the actual trip count of a number of grids is zero, we choose to use absolute metrics (i.e., $MAE$~and~$SAE$) rather than relative metrics to evaluate the inference errors in order to avoid undefined values (i.e. division by zero). ~$R^2$,~$MAE$, and~$SAE$ are calculated by:
\begin{linenomath}
  \begin{align}
  &SS_{tot} = \sum_{i} \left( y_i-\bar{y} \right)^2 \\
  &SS_{res} = \sum_{i} \left( y_i-\hat{y}_i \right)^2 \\
  &R^2 = 1-\frac{SS_{res}}{SS_{tot}} \\
  &MAE = \frac{1}{n}\sum_{i=1}^{n}|y_i-\hat{y}_i|\\
  &SAE = \sum_{i=1}^{n}|y_i-\hat{y}_i|
  \end{align}
\end{linenomath}
where $SS_{tot}$ is total sum of squares; $SS_{res}$ is residual sum of squares; $y_i$ is count of actual origins (destinations) for grid $i$; $\bar{y}$ is the mean of $y_i$; $\hat{y}_i$ is count of inferred origins (destinations) for grid $i$.

\begin{figure}[H]
     \centering
     \begin{subfigure}[h]{0.4\textwidth}
         \centering
         \includegraphics[width=\textwidth]{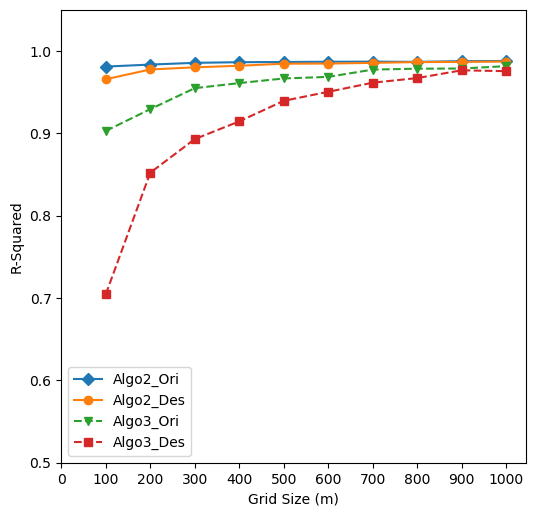}
         \caption{R-Squared}
         \label{fig:r2}
     \end{subfigure}
     \hfill
          \begin{subfigure}[h]{0.4\textwidth}
         \centering
         \includegraphics[width=\textwidth]{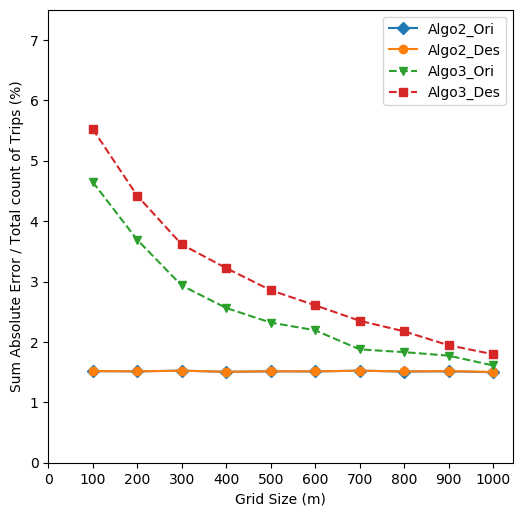}
         \caption{Sum Absolute Error}
         \label{fig:sae}
     \end{subfigure}
     ~
     \begin{subfigure}[h]{0.4\textwidth}
         \centering
         \includegraphics[width=\textwidth]{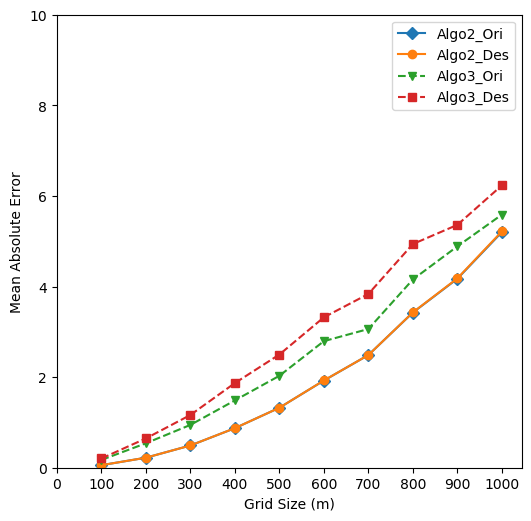}
         \caption{Mean Absolute Error}
         \label{fig:mae}
     \end{subfigure}
     
        \caption{Inference accuracy evaluation of Algorithm~\ref{alg2} and Algorithm~\ref{alg3} for different size of grid (sensitivity analysis)}
        \label{fig:sensitivity}
\end{figure}

To compare the performance of the three inference algorithms, we firstly select a data set with \textit{Static Vehicle ID}. Then we use Algorithm~\ref{alg1} to infer trip origins and destinations as a benchmark. After that, we regenerate vehicle ID using \textit{Resetting Vehicle ID} strategy and \textit{Dynamic Vehicle ID} strategy. We then infer trip origins and destinations using the corresponding algorithms (i.e. Algorithm~\ref{alg2} and Algorithm~\ref{alg3}) and compare the results with the benchmark. Among the six vendors, Spin uses \textit{Static Vehicle ID} and updates the data feeds in a high frequency. Therefore, we use data feeds published by Spin for our algorithm evaluation. To analyze sensitivity to zone size, we calculate the fit metrics for a range of different zone sizes, ranging from 100 m square to 1000 m square zones. The results are presented in Figure~\ref{fig:sensitivity}. Note that the $y$-axis of Figure~\ref{fig:sae} is~$SAE$~divided by total count of trips.

According to Figure~\ref{fig:sensitivity}, both of Algorithm~\ref{alg2} and Algorithm~\ref{alg3} have high inference performance. Algorithm~\ref{alg2} has higher inference accuracy than Algorithm~\ref{alg3}. When the grid size is larger than~$400m \times 400m$, the~$R^2$~of every algorithm is larger than 0.9, and the ~$MAE$~of every algorithm is smaller than 2. From Figure~\ref{fig:sae}, one can notice that as the grid size increases, the~$SAE$~of Algorithm~\ref{alg3} decreases while the~$SAE$~of Algorithm~\ref{alg2} remains flat. As the zones become larger, aggregation leads to a decrease in total error in the system. For example, a $400m\times400m$~area contains sixteen $100m\times100m$~areas, inference errors in each of these sixteen areas may be positive or negative. The positive and negative errors will be neutralized when we calculate the error of the whole $400m\times400m$~area. However, Algorithm~\ref{alg2} already has an extremely high inference accuracy, so the size change of grid has little influence on the corresponding $SAE$. As shown in Figure~\ref{fig:mae}, the~$MAE$~of every algorithm increases as the grid size increases. This is because when the area of a zone becomes larger, both of SAE (numerator of MAE) and the total number of zones (denominator of MAE) decreases, but the denominator decreases much faster than the numerator. Therefore, there is a increasing trend of MAE. In Figure~\ref{fig:mae} and Figure~\ref{fig:sae}, the~$MAE$~is always below 7 and the~$SAE/Total~Count~of~Trips$~is smaller than 6\%. Therefore, we believe the inference error is acceptable for both Algorithm 2 and Algorithm 3.

\begin{figure}[H]
     \centering
     \begin{subfigure}[!h]{1\textwidth}
         \centering
         \includegraphics[width=\textwidth]{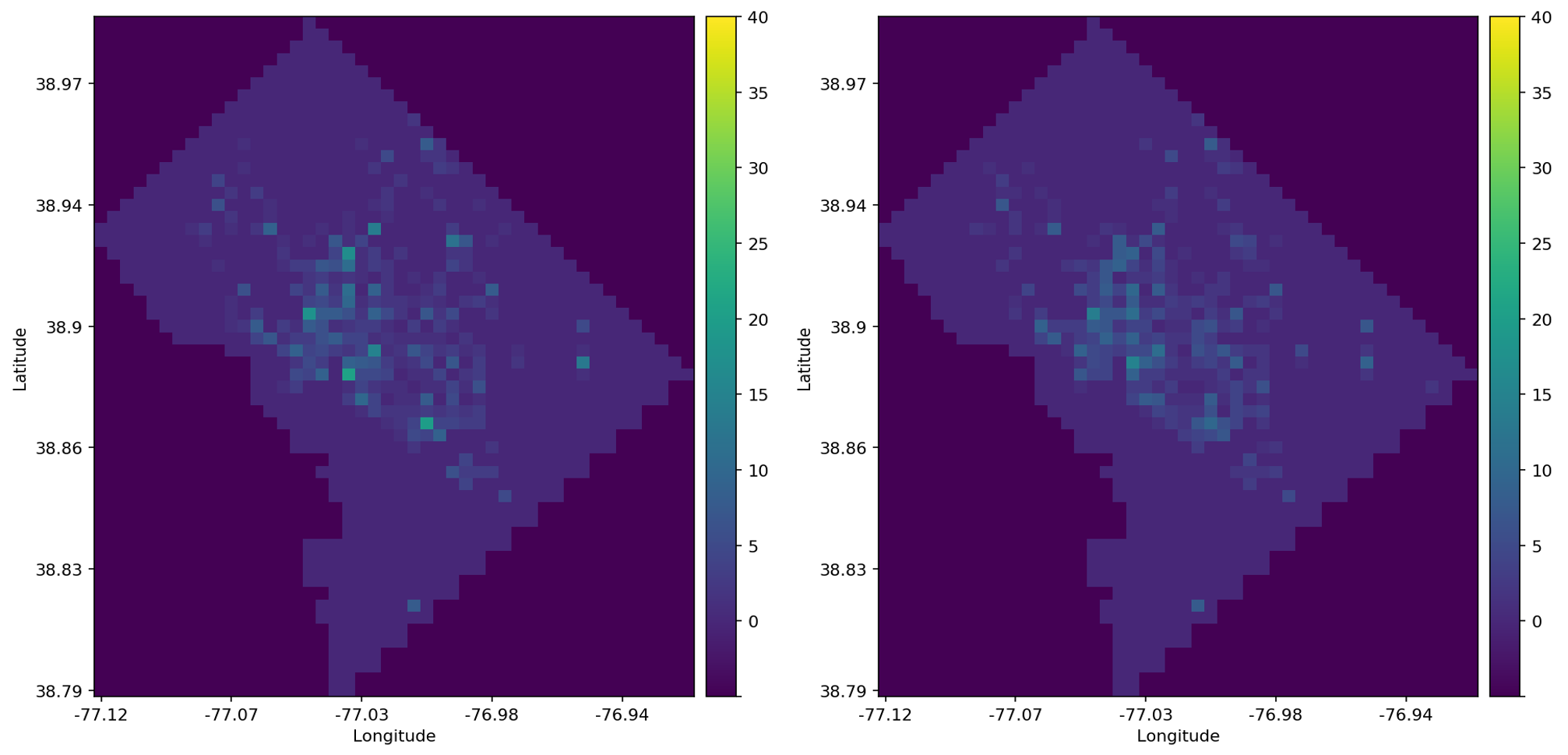}
         \caption{Algorithm~\ref{alg2} (Left: trip origin; Right: trip destination.)}
         \label{fig:erroralg2}
     \end{subfigure}
     \hfill
     \begin{subfigure}[!h]{1\textwidth}
         \centering
         \includegraphics[width=\textwidth]{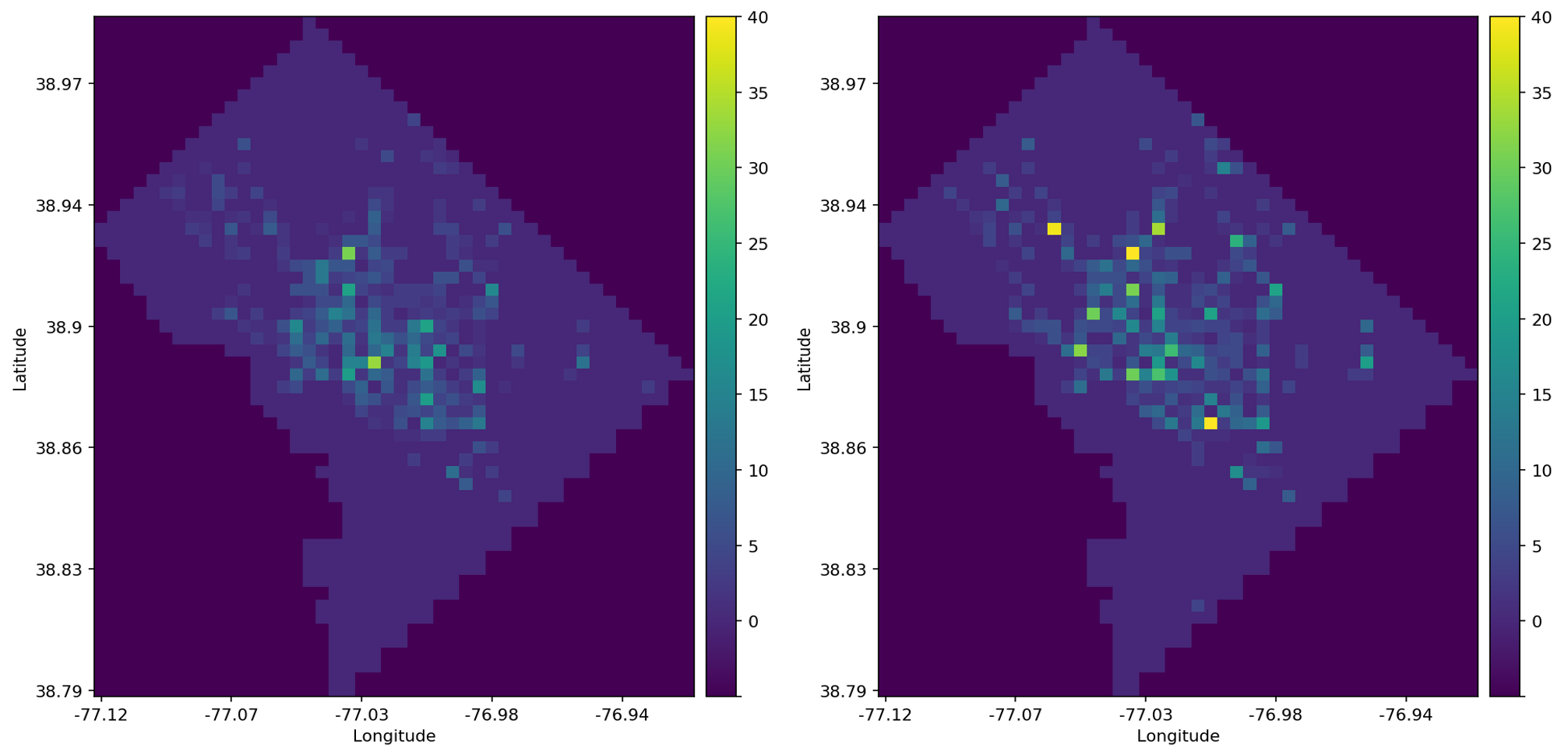}
         \caption{Algorithm~\ref{alg3} (Left: trip origin; Right: trip destination.)}
         \label{fig:erroralg3}
     \end{subfigure}
        \caption{Spatial distributions of inference error (absolute value) of Algorithm~\ref{alg2} and Algorithm~\ref{alg3}}
        \label{fig:error}
\end{figure}

We further examine the spatial distributions of inference error with~$400m\times400m$~grid size. The spatial distributions of inference error of Algorithm~\ref{alg2} and Algorithm~\ref{alg3} are presented in Figure~\ref{fig:error}. We can see that the inference error (absolute value) is larger in downtown areas, where the trip density is greatest. There are more movements of available e-scooters such as vehicle rebalancing in downtown areas. Most launches and eliminations of e-scooters also happen in these areas. As discussed in the methods section, each of these activities may lead to inference errors. Therefore, the inference error is larger in the downtown area.

\section{Discussion and Conclusion}
In this study, we propose a package of algorithms to infer e-scooter origins and destinations from the GBFS data with various vehicle ID types. We examine the accuracy of two proposed algorithms using data from Washington DC. Two algorithms that work with dynamic IDs are tested in comparison to an algorithm that works for static IDs, and is therefore more reliable. We measure inference error with R-squared, mean absolute error, and sum absolute error. With scraped GBFS data from one week, trip origins and destinations are inferred using the proposed algorithms. The inference accuracy of the proposed algorithms are evaluated by simulations based on Spin data (which has static vehicle ID). The results show that both Algorithm 2 and Algorithm 3 have good inference accuracy with ~$R^2$~ larger than 0.9 and ~$MAE$~ smaller than 2 when evaluated with a $400m\times400m$~grid. As the grid size increases, the~$R^2$~of Algorithm~\ref{alg3} increases and the~$SAE$~decreases, while the~$R^2$~and~$SAE$~of Algorithm~\ref{alg2} remains flat. The absolute errors of Algorithm 2 and Algorithm 3 are larger in the downtown areas, where the trip density is highest. 

Using these proposed algorithms, we conduct a temporal analysis on scooter trip origins and destinations. These data show that e-scooter services have a morning peak and an afternoon peak on weekdays, and these peaks are more pronounced compared with the results produced by earlier studies. The findings suggest that e-scooter services may be increasingly used in commuting trips. A spatial analysis with our inferred OD data shows that many areas with high intensity e-scooter usage are close to metro stations. This suggests that travelers are using e-scooter services to connect with transit stations, possibly addressing the ``first mile/last mile'' problem.

Cities have to balance the need for information about transportation systems with the need for traveler privacy. The public transportation authorities require data and information to manage and guide transportation policy in a fast-changing world. For example, public authorities may seek to improve connectivity between micromobility services and the public transit system. However, a growing body of research indicates that ``anonymous'' mobility data, which does not specifically reference a person’s name, email, or address could still be used to re-identify specific individuals, their whereabouts, and activities \citep{regina2019medium,baltra2020data}. For example, information on the starts and ends of trips or the entire GPS trace of a trip can be combined with other datasets to identify specific individuals. Therefore, GPS and location data are considered as privacy-sensitive information in some states \citep[e.g.][]{calaw}. Hence, data standards such as MDS, which track a greater detail of location information, are becoming controversial. MDS requirements have been the subject of some lawsuits over data sharing standards (e.g., Uber and ACLU sued LADOT over MDS \citep{uber2020lawsuit,aclu2020lawsuit}). 
The GBFS data standard, by contrast, has gradually moved away from static vehicle ID to resetting or dynamic vehicle ID. This helps to mitigate privacy concerns. We expect more cities to adopt GBFS rather than MDS standards in the near future.

There is a trade-off between preserving user privacy and extracting useful information from the GBFS data. For GBFS with static vehicle ID, we can directly infer OD pairs; When GBFS is used with dynamic IDs, we can only get unlinked trip origins and destinations for resetting/dynamic vehicle ID. With unlinked ends, it is impossible to gain information about how micromobility trips are being used as a substitute for other modes, which will partially hinder our understanding of micromobility usage patterns. Cities must carefully navigate the balance between privacy protection and data utility. If cities failing to protect sensitive data,  e-scooter companies may use such failures as an excuse to claw back data sharing agreements. On the other hand, for transportation officials to promote safe and convenient travel options for all, some mobility data (e.g., OD demand and trip trajectory) are essential to guide policymaking and mobility management. Cities must balance the goal of public data sharing with user privacy, and GBFS seems to currently offer the best balance between these competing objectives. With GBFS with a dynamic vehicle ID,  trip origin and destination information can be inferred by our algorithms; such data may be sufficient for many research and policy setting purposes.

A limitation of this study is that the inference algorithms and data cleaning process require certain assumptions to be made, such as the maximum e-scooter speed and minimum travel distance. Since the trips not made by users such as rebalancing trips are not labeled in GBFS data, we need to use some criteria such as travel distance to exclude these trips. However, these trips cannot necessarily always be eliminated. Another limitation is that the proposed algorithms for GBFS with dynamic IDs can only infer origins and destinations separately (i.e., we cannot link the origin and destination of a trip). Because of the inference error, the count of origins and destinations may not perfectly match. Future work may consider developing optimization algorithms to infer OD pairs from the separate trip origins and destinations \citep{ji2015statistical,ji2015transit}. Moreover, as the number of entities sharing GBFS increases, researches may examine how the GBFS data and the algorithms developed for them such as the ones we present here have been instrumental for transportation decision-making.

\section{Acknowledgements}
This research was supported by the U.S. Department of Transportation through the Southeastern Transportation Research, Innovation, Development and Education (STRIDE) Region 4 University Transportation Center (Grant No. 69A3551747104).






\bibliographystyle{abbrvnat}
\biboptions{semicolon,round,sort,authoryear}
\bibliography{main.bib}

\begin{thebibliography}{18}
\providecommand{\natexlab}[1]{#1}
\providecommand{\url}[1]{\texttt{#1}}
\expandafter\ifx\csname urlstyle\endcsname\relax
  \providecommand{\doi}[1]{doi: #1}\else
  \providecommand{\doi}{doi: \begingroup \urlstyle{rm}\Url}\fi

\bibitem[ACLU(2020 (Accessed July 30, 2020))]{aclu2020lawsuit}
ACLU.
\newblock Privacy lawsuit your scooter gps data being tracked.
\newblock \emph{ACLU}, 2020 (Accessed July 30, 2020).
\newblock URL
  \url{https://www.aclusocal.org/en/press-releases/privacy-lawsuit-your-scooter-gps-data-being-tracked}.

\bibitem[Bai and Jiao(2020)]{bai2020dockless}
S.~Bai and J.~Jiao.
\newblock Dockless e-scooter usage patterns and urban built environments: a
  comparison study of austin, tx, and minneapolis, mn.
\newblock \emph{Travel Behaviour and Society}, 20:\penalty0 264--272, 2020.

\bibitem[Baltra et~al.(2020)Baltra, Imana, Jiang, and Korolova]{baltra2020data}
G.~Baltra, B.~Imana, W.~Jiang, and A.~Korolova.
\newblock On the data fight between cities and mobility providers.
\newblock \emph{arXiv preprint arXiv:2004.09072}, 2020.

\bibitem[{California Law}(1967)]{calaw}
{California Law}.
\newblock California law part 1. of crime and punishments, title 15.
  miscellaneous crimes, chapter 1.5. invasion of privacy.
\newblock \emph{California Law}, 1967.

\bibitem[Caspi et~al.(2020)Caspi, Smart, and Noland]{caspi2020spatial}
O.~Caspi, M.~J. Smart, and R.~B. Noland.
\newblock Spatial associations of dockless shared e-scooter usage.
\newblock \emph{Transportation Research Part D: Transport and Environment},
  86:\penalty0 102396, 2020.

\bibitem[{City of Austin}(2020 (Accessed July 6, 2020))]{austin2020trip}
{City of Austin}.
\newblock Shared micromobility vehicle trips.
\newblock \emph{TX}, 2020 (Accessed July 6, 2020).
\newblock URL
  \url{https://data.austintexas.gov/Transportation-and-Mobility/Shared-Micromobility-Vehicle-Trips/7d8e-dm7r}.

\bibitem[{City of Minneapolis}(2020 (Accessed July 6, 2020))]{minn2020trip}
{City of Minneapolis}.
\newblock Motorized foot scooter trips 2018.
\newblock \emph{MN}, 2020 (Accessed July 6, 2020).
\newblock URL
  \url{http://opendata.minneapolismn.gov/datasets/motorized-foot-scooter-trips-2018}.

\bibitem[Clewlow(2019 (Accessed July 30, 2020))]{regina2019medium}
R.~Clewlow.
\newblock Finding the right balance between mobility data-sharing in cities and
  personal privacy.
\newblock \emph{Medium}, 2019 (Accessed July 30, 2020).
\newblock URL
  \url{https://medium.com/populus-ai/finding-the-right-balance-between-mobility-data-sharing-in-cities-and-personal-privacy-78d941d07908}.

\bibitem[Ji et~al.(2015{\natexlab{a}})Ji, Mishalani, and McCord]{ji2015transit}
Y.~Ji, R.~G. Mishalani, and M.~R. McCord.
\newblock Transit passenger origin--destination flow estimation: Efficiently
  combining onboard survey and large automatic passenger count datasets.
\newblock \emph{Transportation Research Part C: Emerging Technologies},
  58:\penalty0 178--192, 2015{\natexlab{a}}.

\bibitem[Ji et~al.(2015{\natexlab{b}})Ji, You, Jiang, and
  Zhang]{ji2015statistical}
Y.~Ji, Q.~You, S.~Jiang, and H.~M. Zhang.
\newblock Statistical inference on transit route-level origin--destination
  flows using automatic passenger counter data.
\newblock \emph{Journal of Advanced Transportation}, 49\penalty0 (6):\penalty0
  724--737, 2015{\natexlab{b}}.

\bibitem[Liu et~al.(2019)Liu, Seeder, Li, et~al.]{liu2019analysis}
M.~Liu, S.~Seeder, H.~Li, et~al.
\newblock Analysis of e-scooter trips and their temporal usage patterns.
\newblock \emph{Institute of Transportation Engineers. ITE Journal},
  89\penalty0 (6):\penalty0 44--49, 2019.

\bibitem[McKenzie(2019)]{mckenzie2019spatiotemporal}
G.~McKenzie.
\newblock Spatiotemporal comparative analysis of scooter-share and bike-share
  usage patterns in washington, dc.
\newblock \emph{Journal of transport geography}, 78:\penalty0 19--28, 2019.

\bibitem[McKenzie(2020)]{mckenzie2020urban}
G.~McKenzie.
\newblock Urban mobility in the sharing economy: A spatiotemporal comparison of
  shared mobility services.
\newblock \emph{Computers, Environment and Urban Systems}, 79:\penalty0 101418,
  2020.

\bibitem[{NACTO}(2019)]{nacto2019a}
{NACTO}.
\newblock Shared micromobility in the u.s.: 2018.
\newblock \emph{{New York, NY}}, 2019.

\bibitem[{North American Bikeshare Association}(2020 (Accessed July 6,
  2020))]{nabsa2020github}
{North American Bikeshare Association}.
\newblock General bikeshare feed specification.
\newblock \emph{NABSA}, 2020 (Accessed July 6, 2020).
\newblock URL \url{https://github.com/NABSA/gbfs}.

\bibitem[Teale(2020 (Accessed July 30, 2020))]{uber2020lawsuit}
C.~Teale.
\newblock Uber sues ladot over data-sharing requirements.
\newblock \emph{Smart Cities Dive}, 2020 (Accessed July 30, 2020).
\newblock URL
  \url{https://www.smartcitiesdive.com/news/uber-jump-sues-los-angeles-mobility-data-sharing-requirement/574893}.

\bibitem[Zhu et~al.(2020)Zhu, Zhang, Kondor, Santi, and
  Ratti]{zhu2020understanding}
R.~Zhu, X.~Zhang, D.~Kondor, P.~Santi, and C.~Ratti.
\newblock Understanding spatio-temporal heterogeneity of bike-sharing and
  scooter-sharing mobility.
\newblock \emph{Computers, Environment and Urban Systems}, 81:\penalty0 101483,
  2020.

\bibitem[Zou et~al.(2020)Zou, Younes, Erdo{\u{g}}an, and
  Wu]{zou2020exploratory}
Z.~Zou, H.~Younes, S.~Erdo{\u{g}}an, and J.~Wu.
\newblock Exploratory analysis of real-time e-scooter trip data in washington,
  dc.
\newblock \emph{Transportation Research Record}, page 0361198120919760, 2020.

\end{thebibliography}







\end{document}